\documentclass[journal]{IEEEtran}
\makeatletter\if@twocolumn\PassOptionsToPackage{switch}{lineno}\else\fi\makeatother

\usepackage{graphicx}
\usepackage{xurl}
\usepackage[T1]{fontenc}
\ifCLASSINFOpdf
\else
\fi
\usepackage{amsmath}
\usepackage[cmintegrals]{newtxmath}
\usepackage{url,multirow,morefloats,floatflt,cancel,tfrupee}
\makeatletter

\AtBeginDocument{\@ifpackageloaded{textcomp}{}{\usepackage{textcomp}}}
\makeatother
\usepackage{colortbl}
\usepackage{xcolor}
\usepackage{pifont}
\usepackage[nointegrals]{wasysym}
\makeatletter

\def\mcWidth#1{\csname TY@F#1\endcsname+\tabcolsep}

\def\cAlignHack{\rightskip\@flushglue\leftskip\@flushglue\parindent\z@\parfillskip\z@skip}
\def\rAlignHack{\rightskip\z@skip\leftskip\@flushglue \parindent\z@\parfillskip\z@skip}

\@ifundefined{etal}{}{}

\usepackage{ifxetex}
\ifxetex\else\if@twocolumn\@ifpackageloaded{stfloats}{}{\usepackage{dblfloatfix}}\fi\fi

\AtBeginDocument{
\expandafter\ifx\csname eqalign\endcsname\relax
\def\eqalign#1{\null\vcenter{\def\\{\cr}\openup\jot\m@th
  \ialign{\strut$\displaystyle{##}$\hfil&$\displaystyle{{}##}$\hfil
      \crcr#1\crcr}}\,}
\fi
}

\AtBeginDocument{%
  \@ifpackageloaded{endfloat}%
   {\renewcommand\efloat@iwrite[1]{\immediate\expandafter\protected@write\csname efloat@post#1\endcsname{}}}{\newif\ifefloat@tables}%
}%

\def\BreakURLText#1{\@tfor\brk@tempa:=#1\do{\brk@tempa\hskip0pt}}
\let\lt=<
\let\gt=>
\def\processVert{\ifmmode|\else\textbar\fi}

\@ifundefined{subparagraph}{
\def\subparagraph{\@startsection{paragraph}{5}{2\parindent}{0ex plus 0.1ex minus 0.1ex}%
{0ex}{\normalfont\small\itshape}}%
}{}

\newcommand\role[1]{\unskip}
\newcommand\aucollab[1]{\unskip}
  
\@ifundefined{tsGraphicsScaleX}{\gdef\tsGraphicsScaleX{1}}{}
\@ifundefined{tsGraphicsScaleY}{\gdef\tsGraphicsScaleY{.9}}{}
\def\checkGraphicsWidth{\ifdim\Gin@nat@width>\linewidth
	\tsGraphicsScaleX\linewidth\else\Gin@nat@width\fi}

\def\checkGraphicsHeight{\ifdim\Gin@nat@height>.9\textheight
	\tsGraphicsScaleY\textheight\else\Gin@nat@height\fi}

\def\fixFloatSize#1{}
\let\ts@includegraphics\includegraphics

\def\inlinegraphic[#1]#2{{\edef\@tempa{#1}\edef\baseline@shift{\ifx\@tempa\@empty0\else#1\fi}\edef\tempZ{\the\numexpr(\numexpr(\baseline@shift*\f@size/100))}\protect\raisebox{\tempZ pt}{\ts@includegraphics{#2}}}}

\AtBeginDocument{\def\includegraphics{\@ifnextchar[{\ts@includegraphics}{\ts@includegraphics[width=\checkGraphicsWidth,height=\checkGraphicsHeight,keepaspectratio]}}}

\DeclareMathAlphabet{\mathpzc}{OT1}{pzc}{m}{it}

\def\URL#1#2{\@ifundefined{href}{#2}{\href{#1}{#2}}}

\def\UrlOrds{\do\*\do\-\do\~\do\'\do\"\do\-}%
\g@addto@macro{\UrlBreaks}{\UrlOrds}

\edef\fntEncoding{\f@encoding}

\makeatother

\newif\ifmultipleabstract\multipleabstractfalse%
\usepackage{tabulary}
\makeatletter
\AtBeginDocument{\@ifpackageloaded{longtable}{%
\def\LT@makecaption#1#2#3{%
  \LT@mcol\LT@cols c{\hbox to\z@{\hss\parbox[t]\LTcapwidth{%
    \sbox\@tempboxa{#1{#2: } #3}%
    \ifdim\wd\@tempboxa>\hsize
      #1{#2: }\textsc{#3}%
    \else
      \hbox to\hsize{\hfil\box\@tempboxa\hfil}%
    \fi
    \endgraf\vskip\baselineskip}%
  \hss}}}
}{}}
\makeatother

\makeatletter
\let\citep\cite
\let\citet\cite
\makeatother

   \makeatletter
  \def\fig@textbf{\textbf}
   \AtBeginDocument{\renewcommand\floatc@plain[2]{\setbox\@tempboxa\hbox{{\footnotesize#1.}\footnotesize\hskip.5em#2}%
    \ifdim\wd\@tempboxa>\hsize {\fig@textbf{\footnotesize#1.}}\footnotesize\hskip.5em#2\par
        \else\hbox to\hsize{\hfil\box\@tempboxa\hfil}\fi}}
    \makeatother

\usepackage{float}

\begin{document}

%

        \title{Empirical Study of Straggler Problem in Parameter Server on Iterative Convergent Distributed Machine Learning}
      
\author{Benjamin~Wong\thanks{Benjamin~Wong is with Computer Science, Chinese University of Hong Kong, e-mail: tszkan.wong@link.cuhk.edu.hk (Corresponding author).}}

\maketitle 

\begin{abstract}
The purpose of this study is to test the effectiveness of  current straggler mitigation techniques over different important iterative convergent machine learning(ML) algorithm including Matrix Factorization (MF), Multinomial\textbf{\space }Logistic Regression (MLR), and Latent Dirichlet Allocation (LDA\textbf{)}\textbf{\space }. The experiment was conducted to implemented using the FlexPS system, which is the latest system implementation that employ \textbf{parameter server architecture} \textbf{\space }\unskip~\cite{2011140:28554698}, \unskip~\cite{2011140:28554700}. The experiment employed the Bulk Synchronous Parallel (BSP) computational model to examine the straggler problem in Parameter Server on Iterative Convergent Distributed Machine Learning. Moreover, the current research analyzes the experimental arrangement of the parameter server strategy concerning the parallel learning problems by injecting universal straggler patterns and executing latest mitigation techniques. The findings of the study are significant in that as they will provide the necessary platform for conducting further research into the problem and allow the researcher to compare different methods for various applications. The outcome is therefore expected to facilitate the development of new techniques coupled with new perspectives in addressing this problem. 
\end{abstract}
    
%
\IEEEpeerreviewmaketitle

\section{Introduction}

\subsection{ The Straggler Problem } Different applications employ iterative convergent algorithms, especially machine learnings. Parallel execution of such algorithms is typically expressed through the Bulk Synchronous Parallel (BSP) model for computation \unskip~\cite{2011140:28554699}. Nonetheless, implementation through this model is characterized by what is referred as the straggler problem in computing terminology. \unskip~\cite{2011140:28554693} This means that every brief delay for a given worker can result in a slowdown of the whole process or iteration. This problem significantly reduces the overall completion time and performance, and the problem gets worse with increased parallelism. The diagram below is a visual representation of the straggler problem.

\bgroup
\fixFloatSize{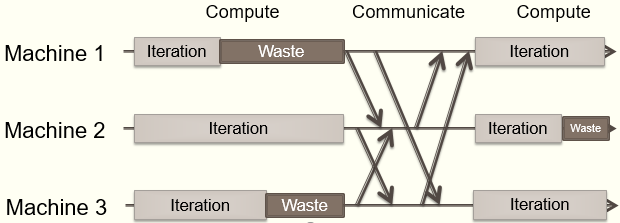}
\begin{figure}[!htbp]
\centering \makeatletter\IfFileExists{images/image1.png}{\includegraphics{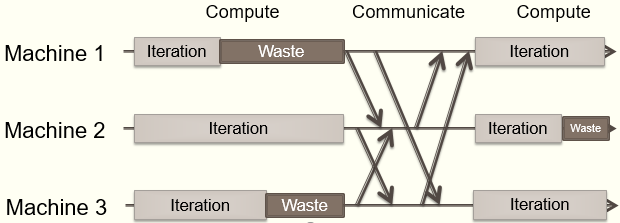}}{\includegraphics{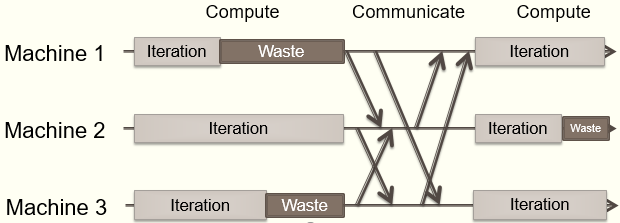}}
\makeatother 
\caption{{The Straggler Problem}}
\label{figure-e9be2216e4314b549ff8a1c2bfc26d6f}
\end{figure}
\egroup
 The training time of a single iteration is estimated depending on the worker with the longest computation time and communication time on each iteration, as expressed in the following: 
\begin{eqnarray*}t_{iteration}=\sum\left(max_{1\leq i\;\leq n\;}Comp\left(i\right)+max_{1\leq i\leq n}Comm(i)\right) \end{eqnarray*}
where $n $ is the number of workers, and $Compute(i) $ and $Communicate(i) $ are computation time and communication time of the worker $i $ respectively.

We can see that any delay in computation or communication of any worker may delay the overall process time. We can further derive the wait time, as expressed in the following: 
\begin{eqnarray*}t_{waste}=\;{\sum_{i=0}^{n}{t_{waste}(i)}} \end{eqnarray*}
Where $n $ is the number of workers, and $t_{waste}(i) $ is the time to wait synchronization barrier of the worker $i $, i.e. wasted time. To address the straggler problem, we have to minimize the total waste time, $t_{waste} $.

\subsection{ Mitigate the Straggler Problem}Stragglers problem have long overwhelmed parallel computing, and numerous technique have been researched to moderate them. Following table summarize current techniques for mitigate straggler problem.

\begin{table}[!htbp]
\caption{{} }
\label{table-wrap-a415f28151464671a767f425340534aa}
\def\arraystretch{1}
\ignorespaces 
\centering 
\begin{tabulary}{\linewidth}{LL}
\hline 
 Mitigation Type &
   Implementations\\
 Speculative execution  &
   ML-NA \unskip~\cite{2011140:28554702}, FlexRR \unskip~\cite{2011140:28554692}\\
 Task Cloning  &
   Wrangler \unskip~\cite{2011140:28554696},Full Clone \unskip~\cite{2011140:28554703}\\
 Job Shedding and Job Stealing &
   FlexRR\\
 Relaxed progress synchronization &
   Stale Synchronization Parallel (SSP) \unskip~\cite{2011140:28554695}\\
\hline 
\end{tabulary}\par 
\end{table}
Summary of Current mitigation techniques

In addressing the straggler problem, new significant data structures have adopted different reactive or preventive speculation approaches through which a system initiates backup or extra copies for tasks on other machines via a prudent manner. Currently, there are 4 popular type of straggler mitigation methods, these including the Speculative execution, task cloning, Job Shedding and Job Stealing and Relaxed Progress Synchronization. 

In the Speculative execution, the development of distinct tasks is examined through the system as well as backup copies when the stragglers are detected. In the cloning approach, on the other hand, extra task copies are organized in a parallel manner with the first job if only the anticipated cost involved in computation is low and the availability of system resources is guaranteed. In the Job Shedding and Job Stealing approach, the system adaptively rebalance work among the workers. In Relaxed progress synchronization approach, the system allow certain level of inconsistency of synchronization and retain the convergence guarantee. 
    
\section{Current Straggler Mitigation Technique}

\subsection{ Speculative execution \& task cloning (Predict Straggler and task cloning) }Research on large scale computing clusters has shown that the completion time of a particular task \unskip~\cite{2011140:28554697}is often unnecessarily delayed by one or a few ``stragglers'' or straggling task that are assigned with either a failing or overloaded server, but definitely there may have a better choice within the cluster. This finding triggered the study of speculative execution so that to diminish stragglers in the processing of data systems like Hadoop, MapReduce, and Spark. Wrangler \unskip~\cite{2011140:28554696} described an approach that predict when the stragglers are going take place and make scheduling decisions to circumvent such situation.

To mitigate stragglers, recent big data framework, as, for example; the MapReduce system or its alternatives have implemented numerous reactive or proactive approaches under which the system will add additional duplicates of a job on other machines in a judicious method. This technique can significantly reduce delays as due to stragglers, as the output from the faster worker can be used without waiting for slower ones, however, with the trade-off of consuming additional resources.

Full cloning of small jobs, without waiting and speculation completely \unskip~\cite{2011140:28554701}. Cloning small jobs increase the machine utilization since workload are small if jobs is small slightly and consumes a small portion of the resources. \unskip~\cite{2011140:28554691}

\subsection{ Using relaxed progress synchronization}In Bulk synchronous parallel (BSP) model, the straggler problem is significant due to the costly barrier synchronization. 

However, the Bulk Synchronous Parallel (BSP) model represents one of the common ways to implement distributed applications of iterative convergent algorithms underlined by an ``input-data-parallel'' methodology. Typically, the model parameter will be stored in the shared data store so that every worker will update throughout the iteration. There exists design of BSP model guarantees the model data will be available to the worker after synchronization instead of immediately available to all worker, and it would enable worker efficiently to use the cached copy. Another design is to ensure the allocation of work to the worker are similar across the iteration, and it would enable the worker to reuse the already loaded data, and thus reduce the overhead of input data movement.

\bgroup
\fixFloatSize{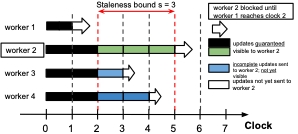}
\begin{figure}[!htbp]
\centering \makeatletter\IfFileExists{images/picturex.png}{\includegraphics{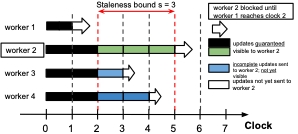}}{\includegraphics{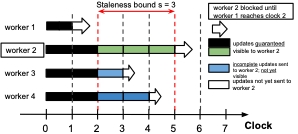}}
\makeatother 
\caption{{Visual representation of Stale Synchronization model}}
\label{figure-e6c8cfebee5e421188e4db52df0cb284}
\end{figure}
\egroup
For some ML Task, strict barriers of BSP can be relaxed with alternate progress synchronization model. Flexible steadiness confines that have been lately proposed suggested and considered two diverse names that are ``Bounded Delay consistency'' and ``Stale Synchronous Parallel'' (SSP) \unskip~\cite{2011140:28554704}. These bonds guarantee convergence proof, as, for example; the SSP has been publicized to moderate insignificant transitory effects of straggler effects. The diagram above shows the visual representation of SSP model. \unskip~\cite{2011140:28554693}

\subsection{ Work Shedding and Work Stealing}Work stealing and shedding of work reflects methodologies for adaptively rebalance work among the workers. The latest research by FlexRR \unskip~\cite{2011140:28554692} proposed a method of work shedding, dedicated to the scope of data-parallel iterative ML. FlexRR employs two mechanisms to mitigate the problem: temporary work reassignment through RapidReassignment (RR), and adjustable consistency parameters through the State Synchronous Parallel (SSP) model. SSP allows individual workers threads to lead the slowest worker by a quantified slack-bound iteration number. This flexibility permits mitigation of the straggler problem to s certain degree but more fundamentally, provides sufficient flexibility to RapidReassignment (RR) making it highly efficient. RR employs peer-to-peer communication thus enabling the self-identification of workers as stragglers. FlexRR designed to use the parameter server method to store and distributing normal states among worker threads. During execution, it will comprise of one process prompting a worker thread for respective cores on the nodes as well as the number of threads in the background for internal functionality.\textbf{\space }Following sequence diagram visualizing the RapidReassignment protocol when a slow worker is behind the progress. 
\bgroup
\fixFloatSize{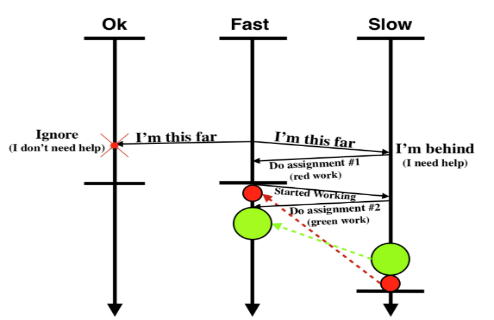}
\begin{figure}[!htbp]
\centering \makeatletter\IfFileExists{images/image3.png}{\includegraphics{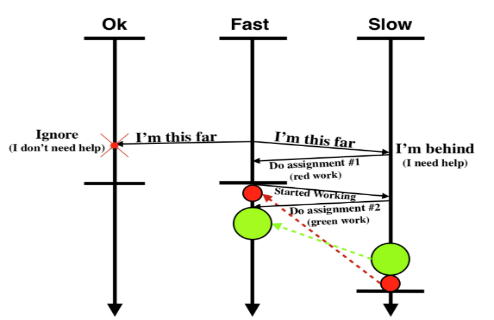}}{\includegraphics{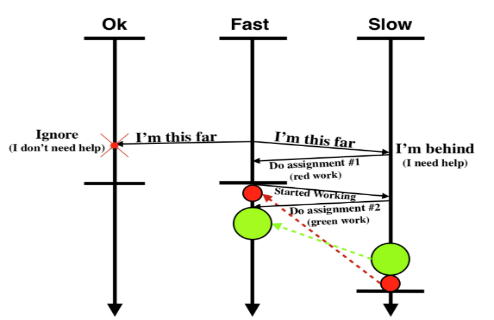}}
\makeatother 
\caption{{Work Shedding and Work Stealing using Rapidreassignment(RR)}}
\label{figure-dd07b2e404f441a1883a659ccf006d16}
\end{figure}
\egroup

\subsection{ Predicting Straggler with ML Technique}Besides Job rescheduling, speculation \unskip~\cite{2011140:28554694} is another standard method for straggler mitigation. It predicts stragglers based on progress score of the task and launch a clone of the job on different machines. Most recent research ML-NA \unskip~\cite{2011140:28554702} is intended on an ML-based Node Performance Analyzer. By leveraging actual task execution log, ML-NA will categorize cluster nodes into different classifications and predict their performance in the following task, so that scheduler can be better and minimize the possibility of task straggler. 
    
\section{Methodology and Experimental Set-Up}
To study the straggler problem, we built a prototype system using FlexPS system, which setup with the Bulk Synchronous Parallel model to support distributed machine learning applications. FlexPS is a ``parameter server'' that provides the abstraction of a set of shared sparse matrices that all processes can access. These matrices are stored in memory, distributed across a set of servers.

FlexPS provides a flexible parallel control API similar to the B{\"{o}}sen system, abstracts consistency management and net- working operations away from application, and presents a simple key-value interface, as shown below table. It provides read and update operations including Get(), Add() and Clock(). 
\begin{table}[!htbp]
\caption{{FlexPS Client API} }
\label{table-wrap-8185957349c247468d7e589815768ec2}
\def\arraystretch{1}
\ignorespaces 
\centering 
\begin{tabulary}{\linewidth}{LL}
\hline 
 Get(key)  &
  Read a parameter indexed by key.\\
 Add(key,value)  &
  Update/Add the parameter key by value\\
 Clock()  &
  Signal end of iteration.\\
\hline 
\end{tabulary}\par 
\end{table}

\subsection{ Injected Straggler Framework}The experiment will test the effects of a straggler beyond what ascended for the duration of the precise instances of the trials. It will also measure attributable delays directly. The research will intentionally inject straggler into the execution, including transient as well as persistent stragglers. The experiment will include different injected frameworks, while the control setup will involve testing the straggler mitigation technique without injected delays. In other words, the control experiment will encompass the Ideal situation representing perfect cases that assume every worker have no wait. The injected straggler pattern are listed in the table below:
\begin{table*}[!htbp]
\caption{{Injected Straggler Framework} }
\label{table-wrap-fa8a05d501174efc9988affb35816f44}
\def\arraystretch{1}
\ignorespaces 
\centering 
\begin{tabulary}{\linewidth}{LL}
\hline 
\textbf{Straggler Pattern} &
  \textbf{Description}\\
 Slow Worker Pattern &
   Random Transient delay, which within iteration job\\
 Disrupted Machine Pattern &
   Transient delay simulated by system disruptor method for specific time\\
 Power Law Pattern &
   The Random Transient delay within iteration job that varies with execution time\\
 Persistent Straggler &
   A particular node that persistently performs slowly \\
 Ideal &
   perfect cases that assume every worker have no wait\\
\hline 
\end{tabulary}\par 
\end{table*}

\subsubsection{ Slow Worker Pattern}Through this pattern, it models transient worker by manually add sleep to the worker, in application or engine. We will add all possibly delay point (any point that may incur waiting) within each iteration, and each worker decide its possibility (x\%) to be delayed independently. It can be delayed from 0 {\textendash} 2 times of the duration of the actual iteration time. It is possible to have no or multiple straggler in each iteration. the transient delay \% will be used to determine the level of worker slowed (e.g., 100\% delay means runs 2 times slower than original one).

The simulation will be implemented by dividing the overall sleep into multiple millisecond sleep, and set to all possibly delay point, (e.g. for a 50\% delay within a 1 second iteration, and there are potential 10 delay points, it can insert a 1ms sleep at each delay point).

\subsubsection{ Disrupted Machine Pattern}Through this methodology approach, the study will model transient resource contention by simulating a high priority disruptor procedure that takes affect significant Memory or CPU resources. It can be duplicated by initiating a disruptor process with a probability of x\% every d seconds. The disruptor process can be implemented by launching multiple threads that work on heavy calculation work (or run infinite loop). This implies that for d seconds, then a temporary delay concentration, which is a component of d, with a machine of p core running p worker threads, the disruptor therefor commences at d \ensuremath{\times} p processes. For instance, 200\% delay on an 8-core machine running eight worker thread will run 16 threads disruptor process \unskip~\cite{2011140:28554698}.

\subsubsection{ Power-Law Pattern}The selection of this methodology was informed by the real-world straggler patterns as established by existing research. The gist of this approach is predicated on the understanding a straggler may occur with a power-law distribution\unskip~\cite{2011140:28554700} to finish an iteration. For instance, 
\begin{eqnarray*}p\;(t)\;\propto\;t\;\times\;\alpha \end{eqnarray*}
Where \ensuremath{\alpha } represents an element of the distribution ``skewness.'' Sleep will, therefore, be employed with the comparable execution in the Slow Worker Pattern, with $t\;\times\;\alpha $ added to the delay time. Lesser $\alpha $ leads to a ``flatter'' distribution. Each iteration employs the same $\alpha $, and without considering the actual completion time of the iteration.

\subsubsection{ Persistent Straggler Pattern}This pattern is caused by a particular node that insistently performs slow. The study also incorporates the concept of persistent straggler pattern whereby some the machines receive 75\% or even 50\% of the work per iteration. Such unbalanced assignments could trigger the case where processing skew of the data is co-related with the data placement.

\subsubsection{ Implementation of Straggler injection framework}Leveraging the FlexPS API, the straggler injection shall be implemented inside the application code. The following operations has implemented to add straggler inside the algorithm implementation. A global configurable switch in the application to control whether current execution will inject straggler simulation.

\begin{table}[!htbp]
\caption{{Straggler injection framework} }
\label{table-wrap-05ecc8dabfbe45af8d6756c164cb9342}
\def\arraystretch{1}
\ignorespaces 
\centering 
\begin{tabulary}{\linewidth}{LL}
\hline 
 InjectStraggler(delay\_percent) &
   Inject Straggler for specific worker, with its delay setting. It will inject delay based on current actual process timer.\\
 CheckPermanentStraggler(worker\_id) \mbox{}\protect\newline  &
   Check whether current worker is elected as permanent straggler\\
 CheckTranisentStraggler(worker\_id, iteration\_count)  \mbox{}\protect\newline  &
  Check whether current worker is elected as transient straggler\\
\hline 
\end{tabulary}\par 
\end{table}
Following Benchmark framework has implemented to extend current FlexPS system in order to measure the performance of the parallel execution. 
\begin{table}[!htbp]
\caption{{Straggler injection framework} }
\label{table-wrap-6dc228372e464f81b0b37dbc80937100}
\def\arraystretch{1}
\ignorespaces 
\centering 
\begin{tabulary}{\linewidth}{L}
\hline 
 Benchmark{\textless}TimeT{\textgreater} A C++ Template Object that encapsulate the benchmark utilities\\
 Benchmark{\textless}TimeT{\textgreater}::measure  Measure the block of code and return the time\_t \\
 Benchmark{\textless}TimeT{\textgreater}::measure\_r  Measure the block of code and return the block's output\\
 Benchmark{\textless}TimeT{\textgreater}::total  Return the total time logged into benchmark object\\
 Benchmark{\textless}TimeT{\textgreater}::reset  Reset the time logged into benchmark object\\
\hline 
\end{tabulary}\par 
\end{table}

\subsection{ Benchmark with selected ML Algorithm }In order to study the behaviour of different mitigation technique over different type of workload, the experiment is designed to conduct using various ML algorithm implemented, and instrument with customized C++ library implemented in FlexPS application code base. Following table describe the dataset used for different ML Algorithm selected in the experiment.
\begin{table*}[!htbp]
\caption{{Experiment dataset used against different algorithm} }
\label{table-wrap-c5f4f5c538dd46de88d38ecb0f5d84fe}
\def\arraystretch{1}
\ignorespaces 
\centering 
\begin{tabulary}{\linewidth}{LL}
\hline Application &  Experiment dataset\\
\hline 
 Logistric Regression (LR) &
   Avazu dataset, containing 40,428,967 training size with a feature dimension of 1,000,000 and 2 classes\\
 Matrix Factorization (MF) &
   Netflix dataset, which is a 480k-by-18k sparse matrix with 100m known features. \\
 Latent Dirichlet Allocation (LDA) &
   New York Times dataset, which is a 1.8 million articles that appeared between January 1, 1987 and June 19, 2007\\
\hline 
\end{tabulary}\par 
\end{table*}

\subsubsection{ Logistic regression (LR)}This is a regression model where DV is unconditional. Specifically, the model employed by this experiment is a binary dependent variable, this means that only two values such as "0" and "1", representing output such as boy/girl, or else win/lose can be used. 

The MLR investigations use the Avazu dataset, containing 40,428,967 training size underlined by a variable dimension of 1,000,000 as well as two classifications. The investigations will also encompass ImageNet datasets coupled with LLC elements comprising 64k experimental observations defined by 1000 classes and feature dimensions of 21,504.

\subsubsection{ Matrix factorization (MF) }This is a procedure employed in reference systems, such as endorsing flicks to consumers in networks such as Netflix. It is also referred to as collaborative filtering. In numerical analysis, given a moderately complete matrix X, matrix factorization factorizes X into factor matrices $L $ and $R $. consequently, their respective product estimates X as $X\;\approx\;LR $. We can, therefore, execute MF through the ``stochastic gradient descent'' (SGD) algorithm with the reference of other systems and individual worker threads. 

The experiment will employ the ``Netflix dataset'' underlined by ``a 480k-by-18k sparse matrix with 100m known features''. These elements are configured in such a way that they factor the data set into the result generated by matrix underlined by rank 100. This version is a ``7683k-by-284l sparse matrix'' characterized by 4.24 billion recognized features underlined by rank 100.

\subsubsection{ Latent Dirichlet Allocation (LDA) }LDA represents a generative probabilistic corpus model. The essence of this ML strategy is predicated on the need to have documents served in random assortments over the possible topic. 

This experiment will employ the Nytimes dataset comprising 100m words within 1.8 million articles containing 100k vocabulary size. They will be configured to categorize words, as well as reports into 1000.

\subsection{Experiment result}We performed experiments on Cluster in University to evaluate all straggler technique in the presence of all straggler patterns injected during the particular times of the experiments.
\begin{table}[!htbp]
\caption{{Experiment result} }
\label{table-wrap-12fb6b24f158478c9ae59c7433d268dd}
\def\arraystretch{1}
\ignorespaces 
\centering 
\begin{tabulary}{\linewidth}{LL}
\hline  Injected Pattern &  Figures\\
\hline 
 Slow worker Pattern &
   Figure 4,5\\
 Disrupted Machine &
   Figure 6\\
 Power Law Pattern &
   Figure 7,8\\
\hline 
\end{tabulary}\par 
\end{table}

\section{Results for Slow Worker pattern}

\bgroup
\fixFloatSize{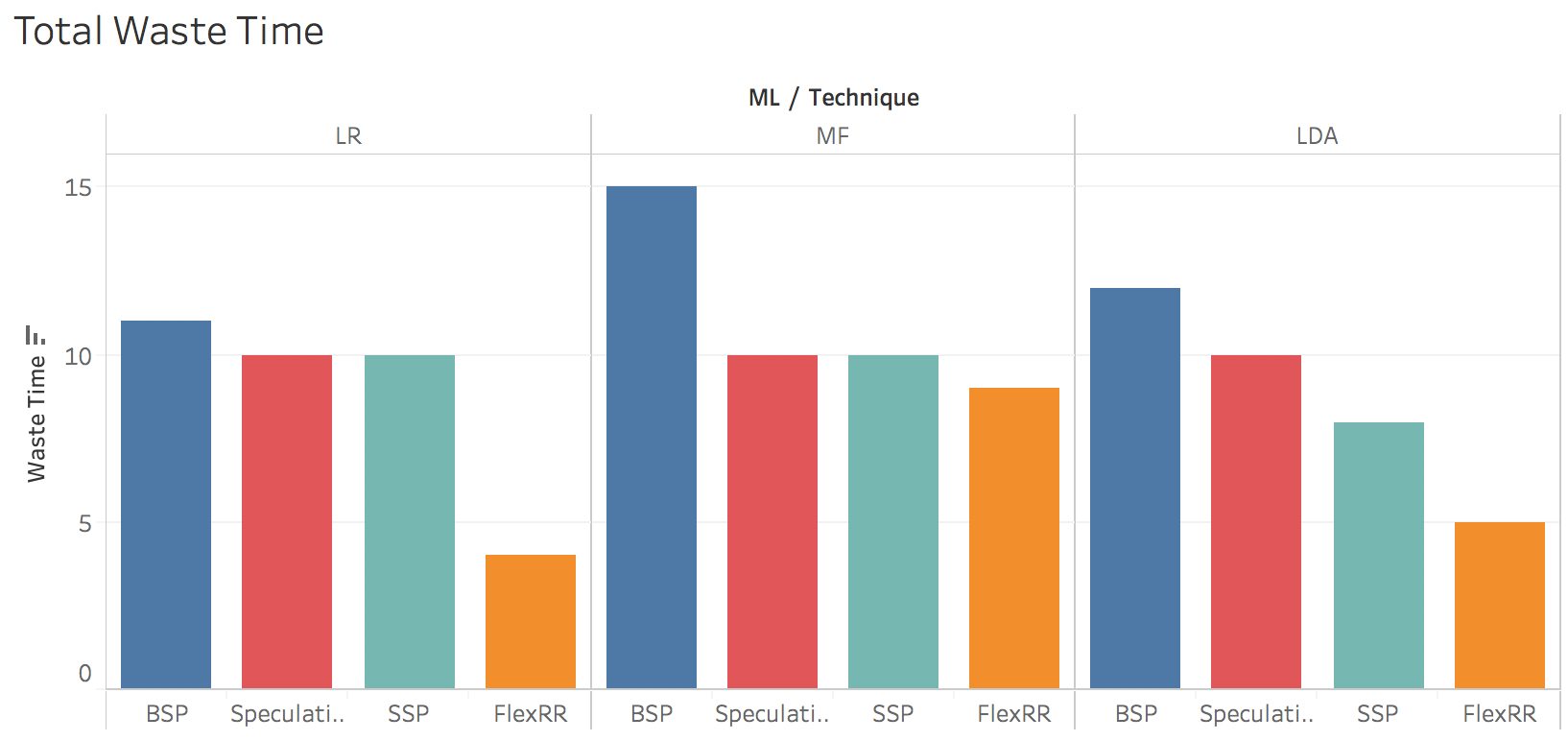}
\begin{figure}[!htbp]
\centering \makeatletter\IfFileExists{images/image4.png}{\includegraphics{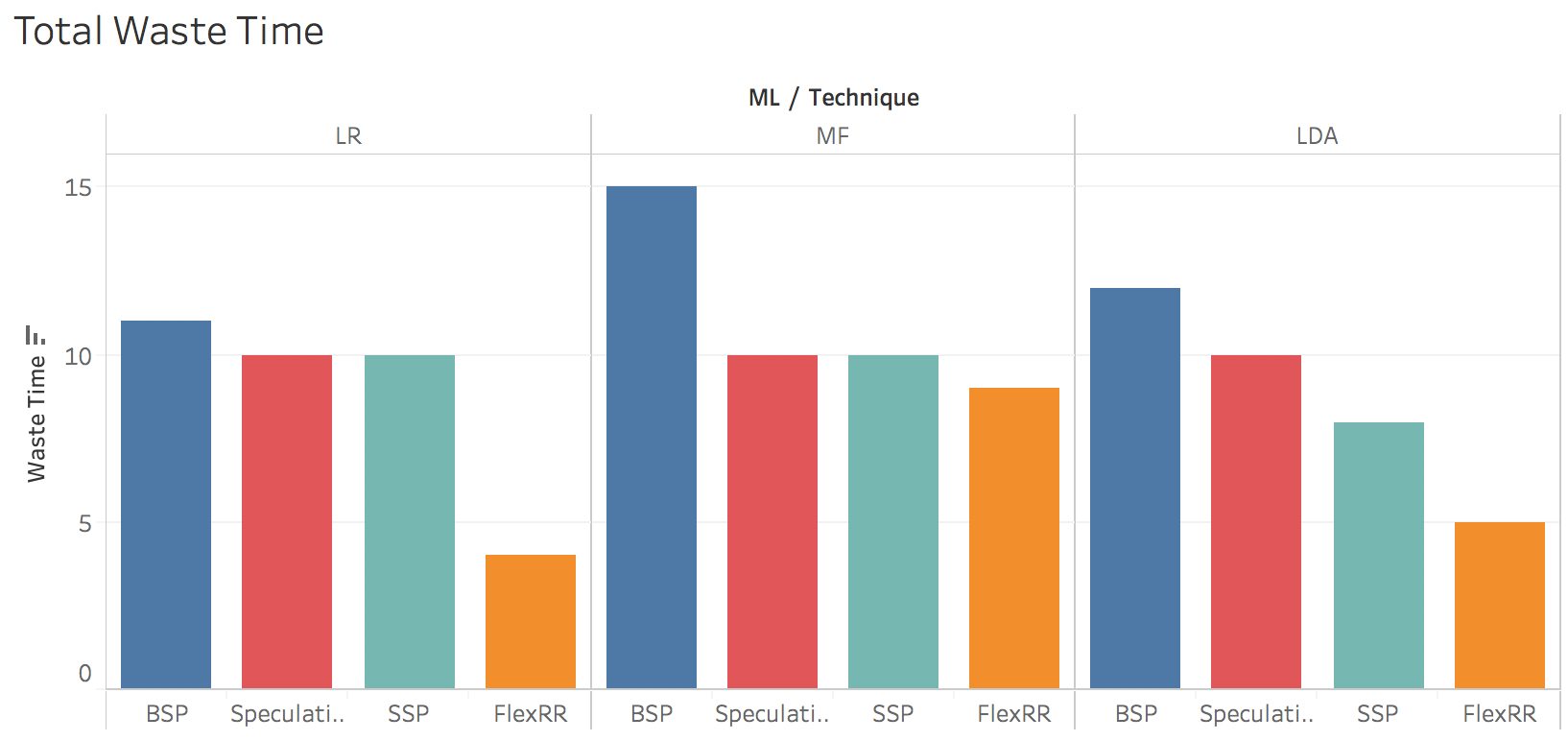}}{\includegraphics{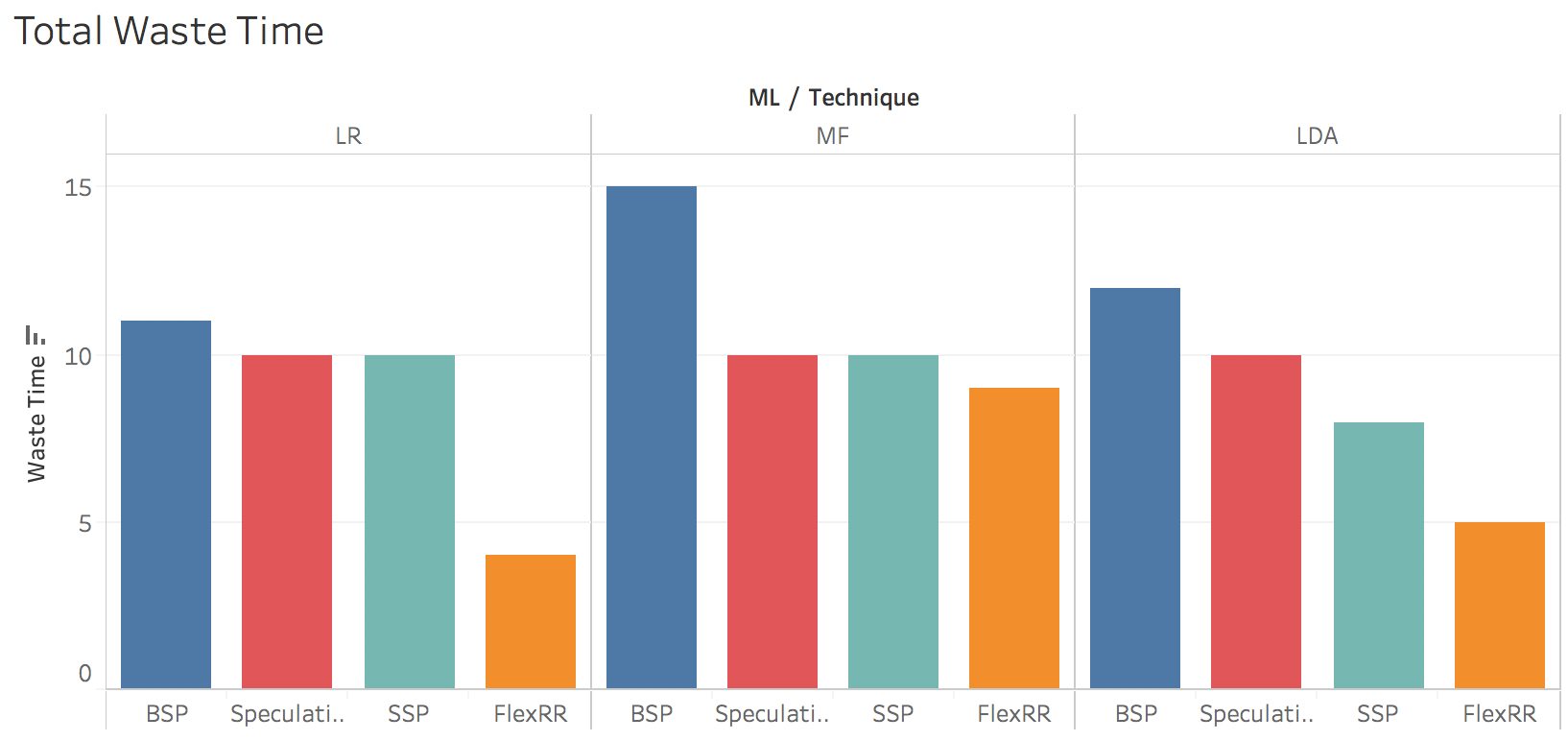}}
\makeatother 
\caption{{Total waste time for Slow worker patter against Technique/ML}}
\label{figure-f4e77419200b4913a8565419c6544775}
\end{figure}
\egroup

\bgroup
\fixFloatSize{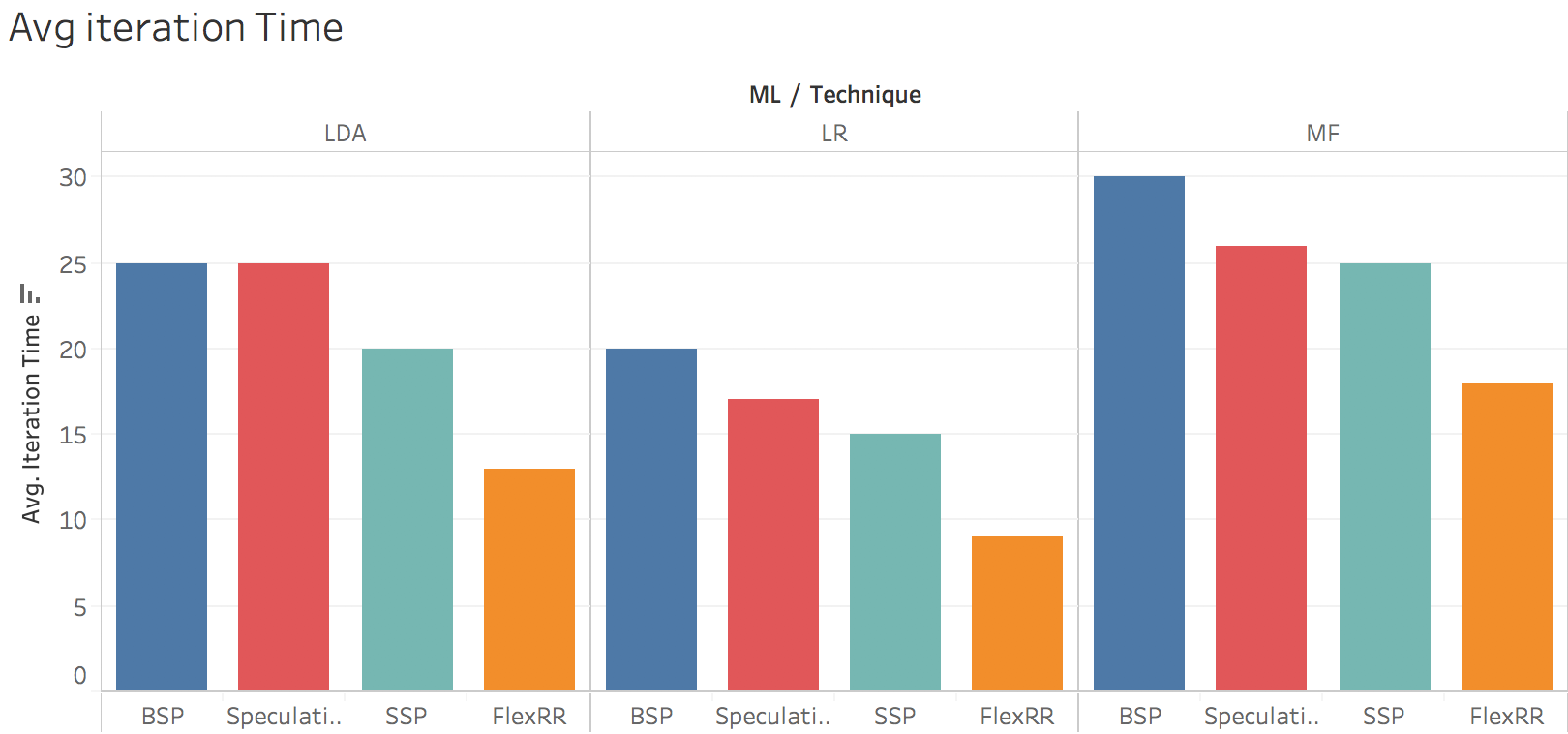}
\begin{figure}[!htbp]
\centering \makeatletter\IfFileExists{images/image5.png}{\includegraphics{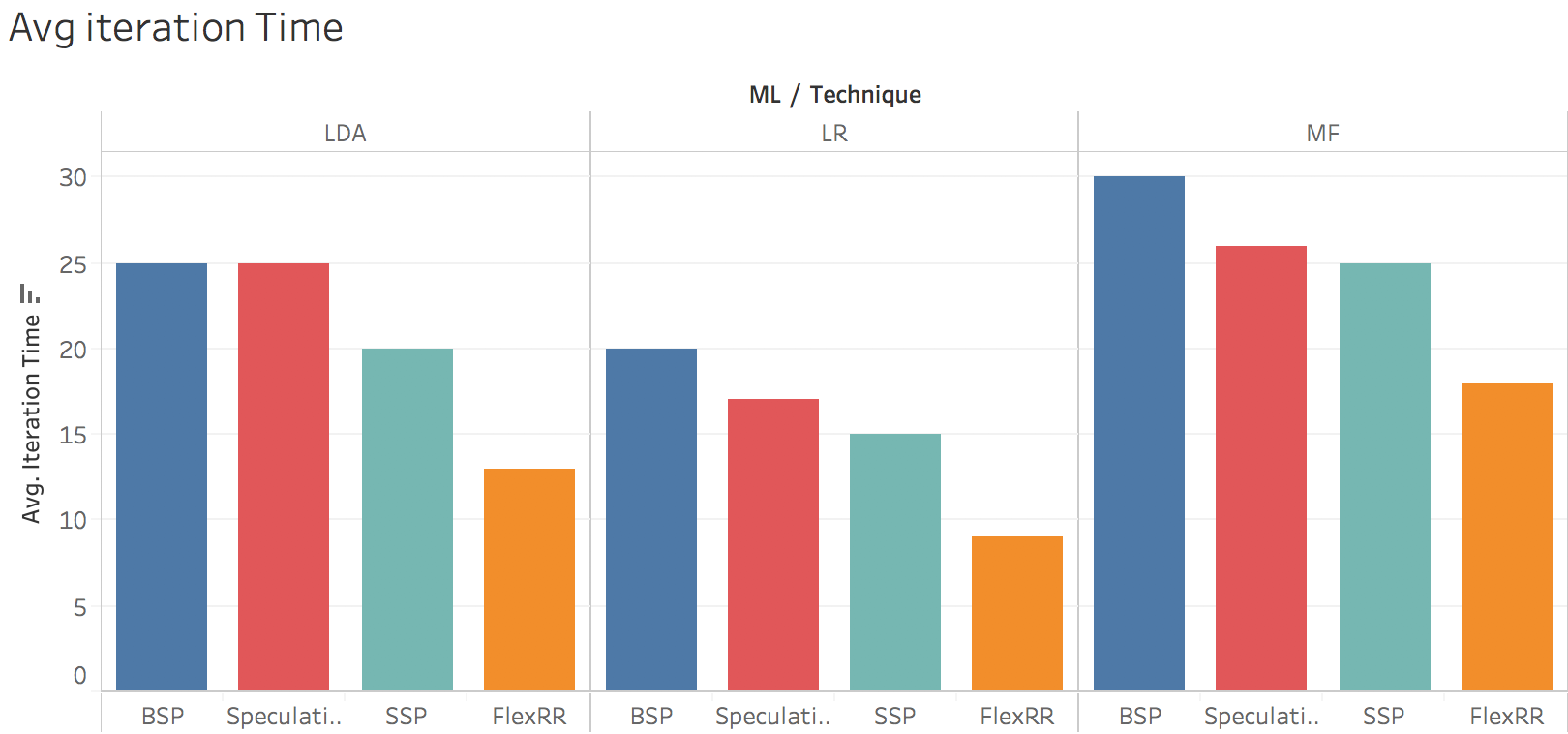}}{\includegraphics{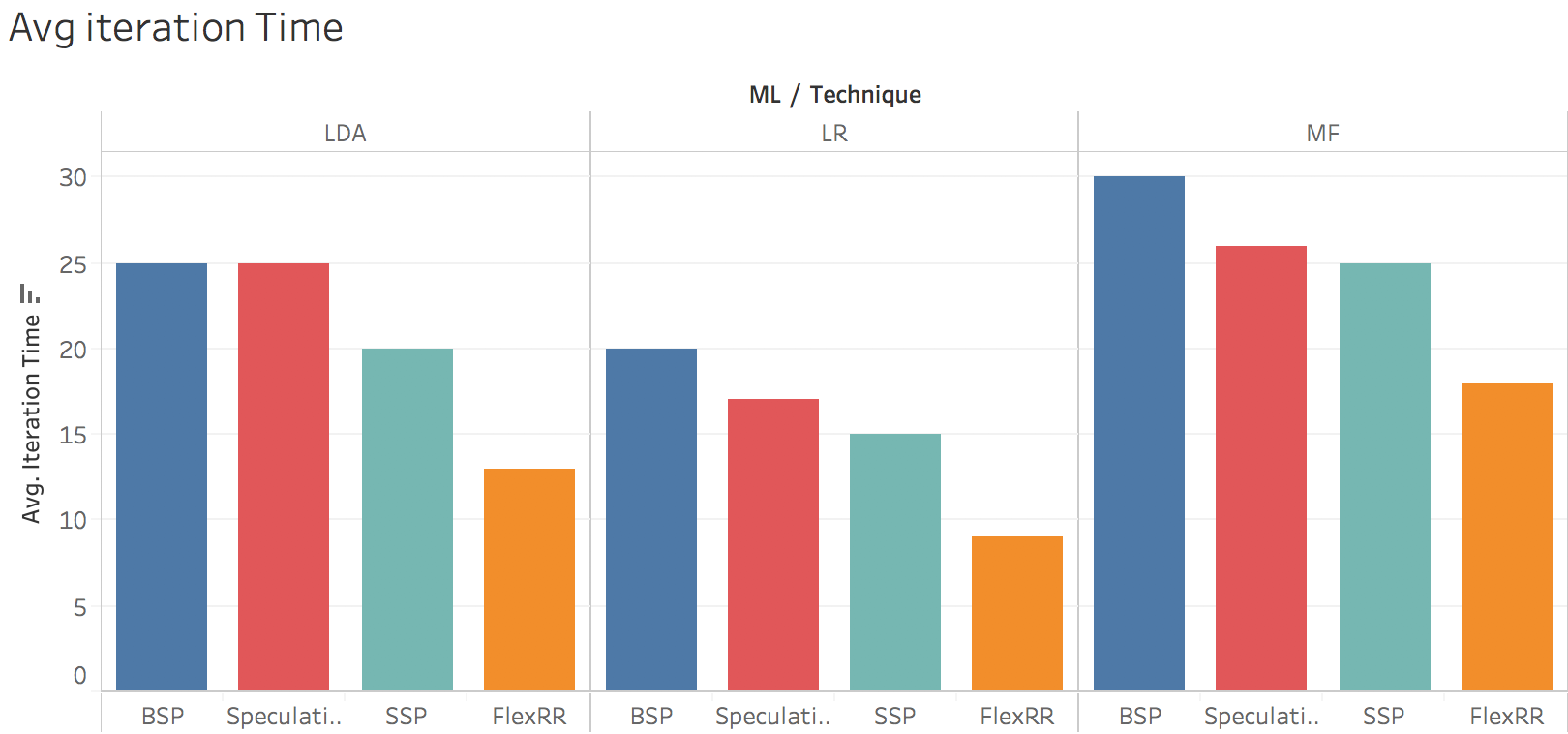}}
\makeatother 
\caption{{Average iteration time for Slow Worker Pattern against Technique/ML}}
\label{figure-d6faa41c0a254ac9a4593f0a09724b5d}
\end{figure}
\egroup
The outcomes of the LR, MF and LDA are presented in figures above. The results show that FlexRR decreases the time-per-iteration or total waste time by 35\% about Matrix factorization\textbf{\space }and 34\% for Latent Dirichlet Allocation relative to the Bulk Synchronous Parallel (BSP) and 25\% and 34\% reductions relative to SSP respectively. As noted in the two experiments, the increase is more significant for LR since the less costly LR undergo more transient effects of the straggler problem.

~The enhancements in cluster were observed despite implementing comparatively short experiments on relatively costly situations expected minimal exhibit sharing of resources with other tenant actions while at the same time highlighting the realness of transient stragglers within infrastructures. 
    
\section{Result for Disrupted Machine Pattern}

\bgroup
\fixFloatSize{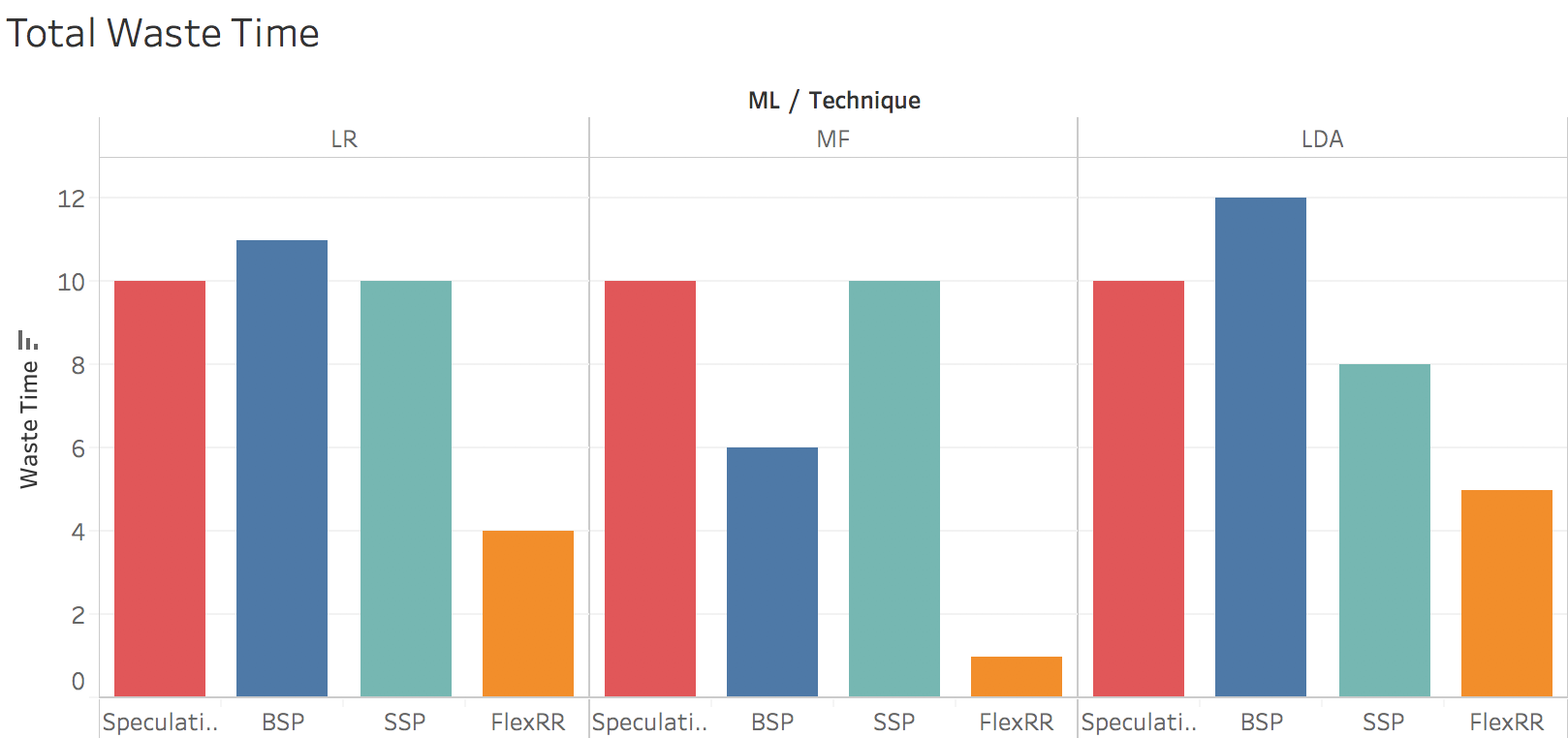}
\begin{figure}[!htbp]
\centering \makeatletter\IfFileExists{images/image6.png}{\includegraphics{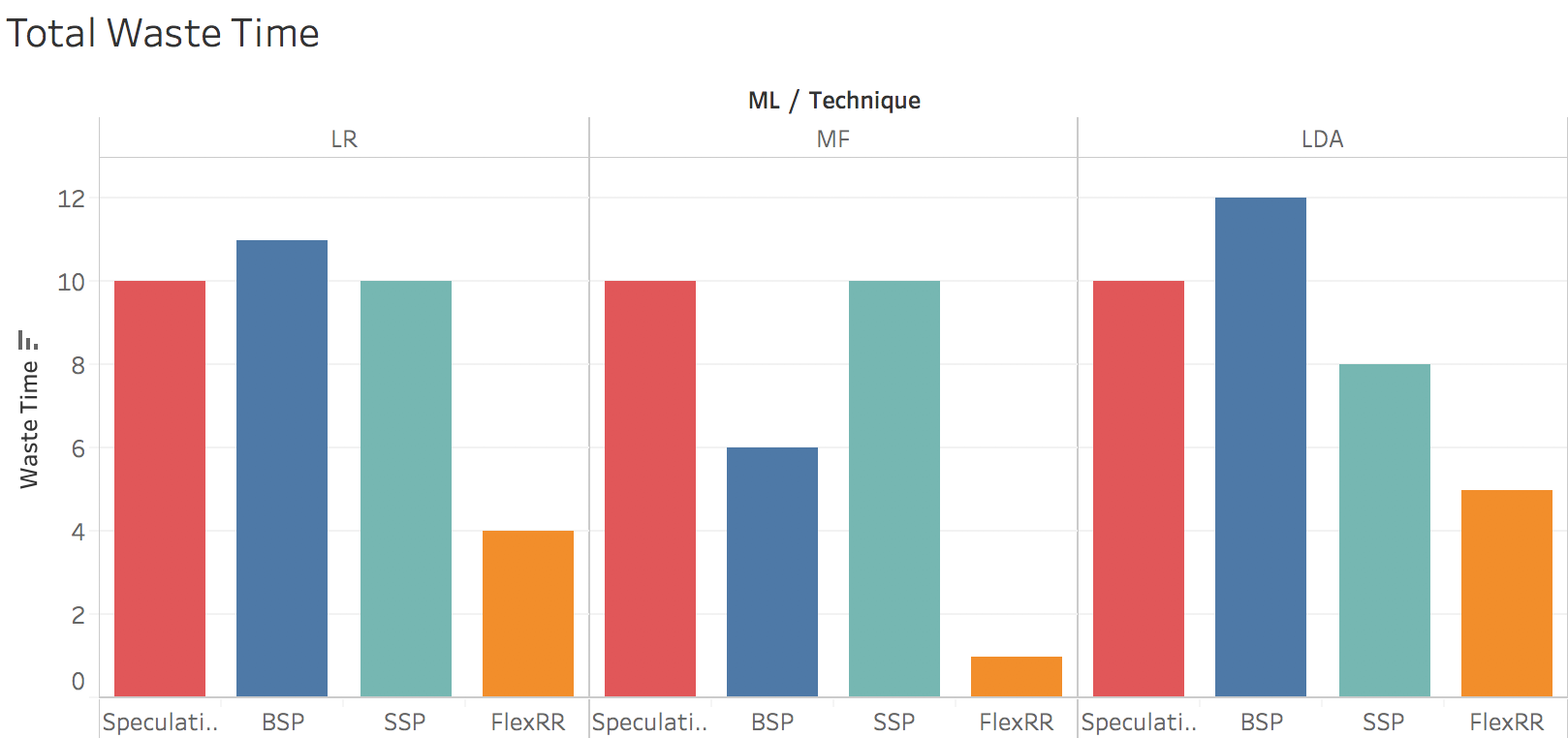}}{\includegraphics{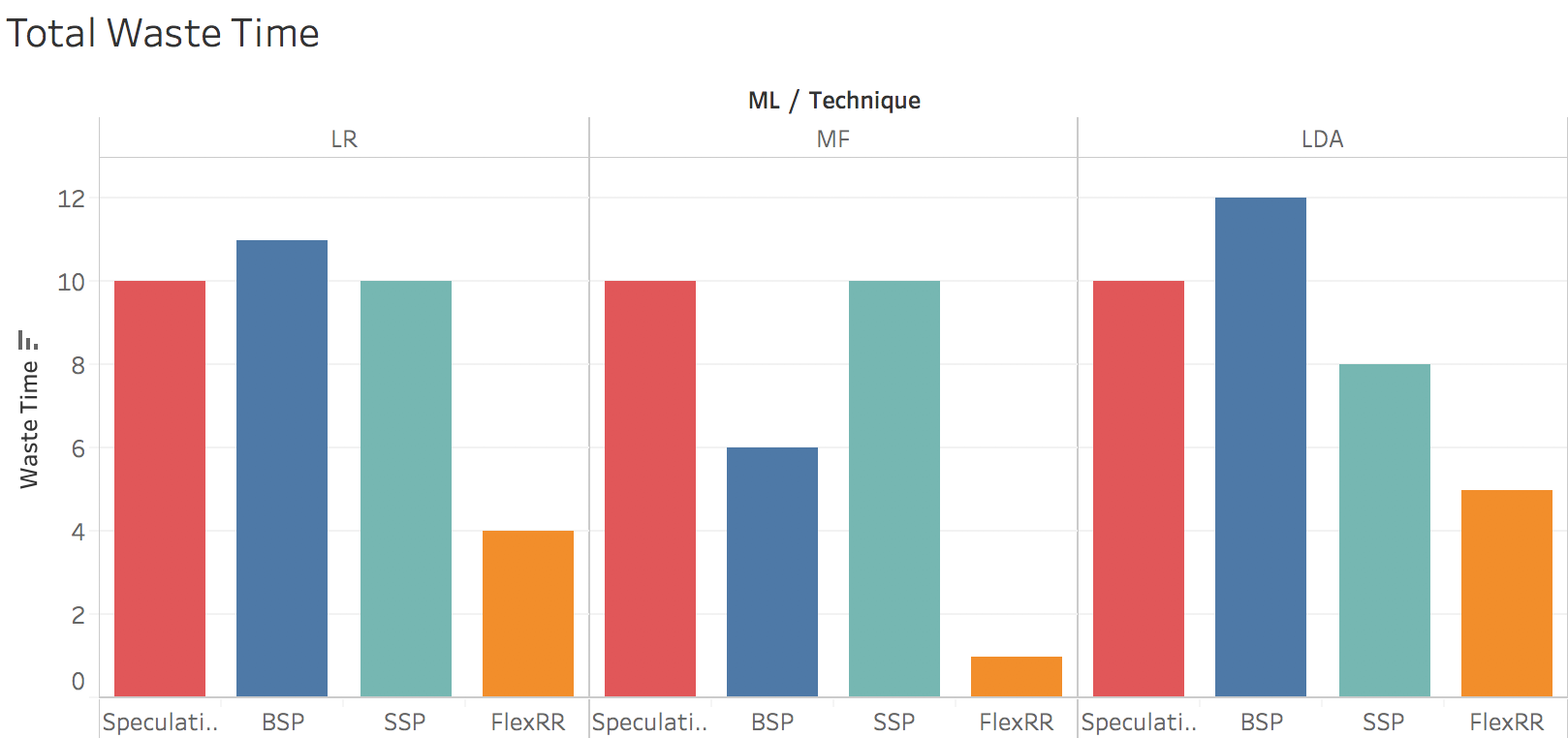}}
\makeatother 
\caption{{Total Waste time for Disrupted machine pattern}}
\label{figure-92901406c0344c6da75abdc3ebb950e0}
\end{figure}
\egroup
~~~We compare the average time-per-iteration (20 iterations) and total waste time of alternative modes for the Disrupted Machine Pattern. Following diagram shows results for MF{\textemdash}results for LDA and MLR are qualitatively similar. SSP and BSP RR individually reduce the delay experienced by BSP by up to 49\% and 42\%, respectively. The combination of the two techniques in FlexRR matches Ideal, reducing the run-time by up to 63\%, very close to ideal waste time.
    
\section{Result for Power Law Machine Pattern}

\bgroup
\fixFloatSize{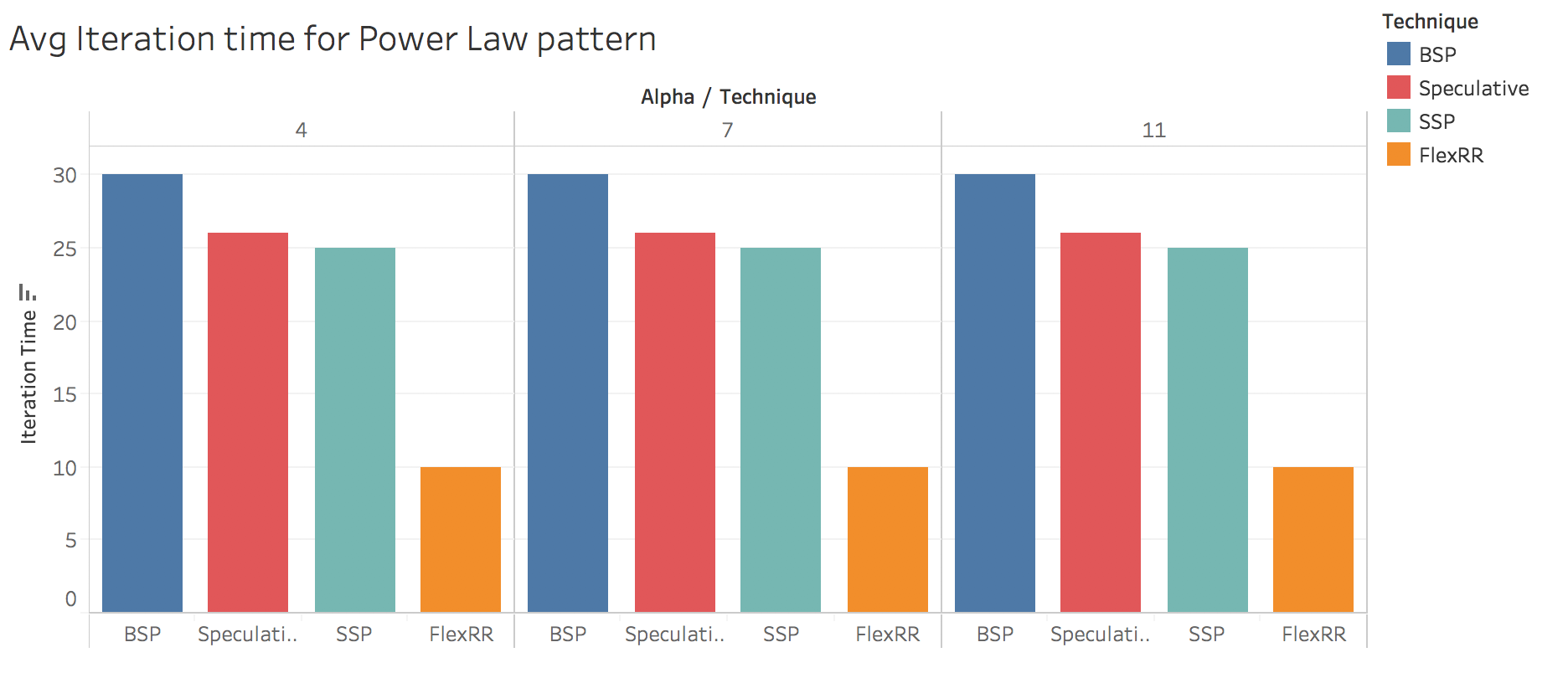}
\begin{figure}[!htbp]
\centering \makeatletter\IfFileExists{images/image7.png}{\includegraphics{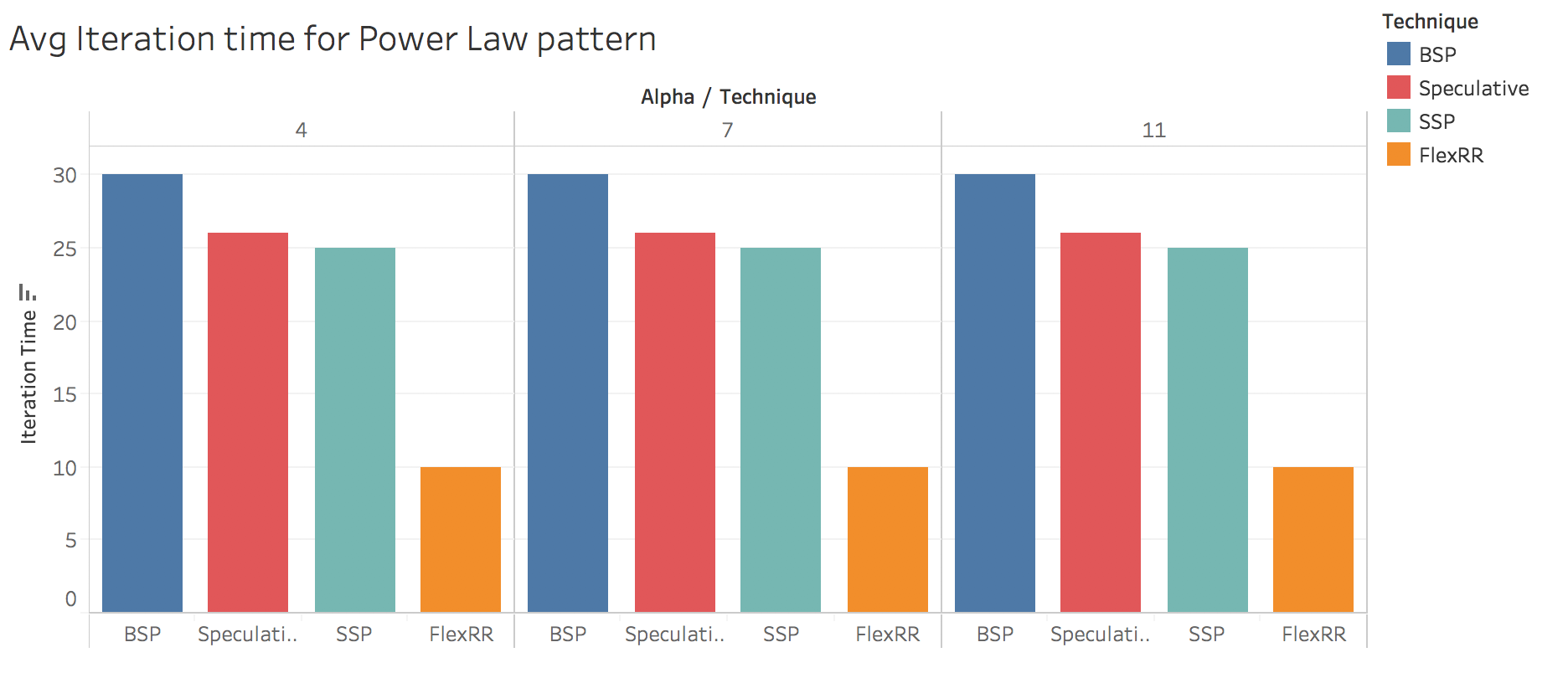}}{\includegraphics{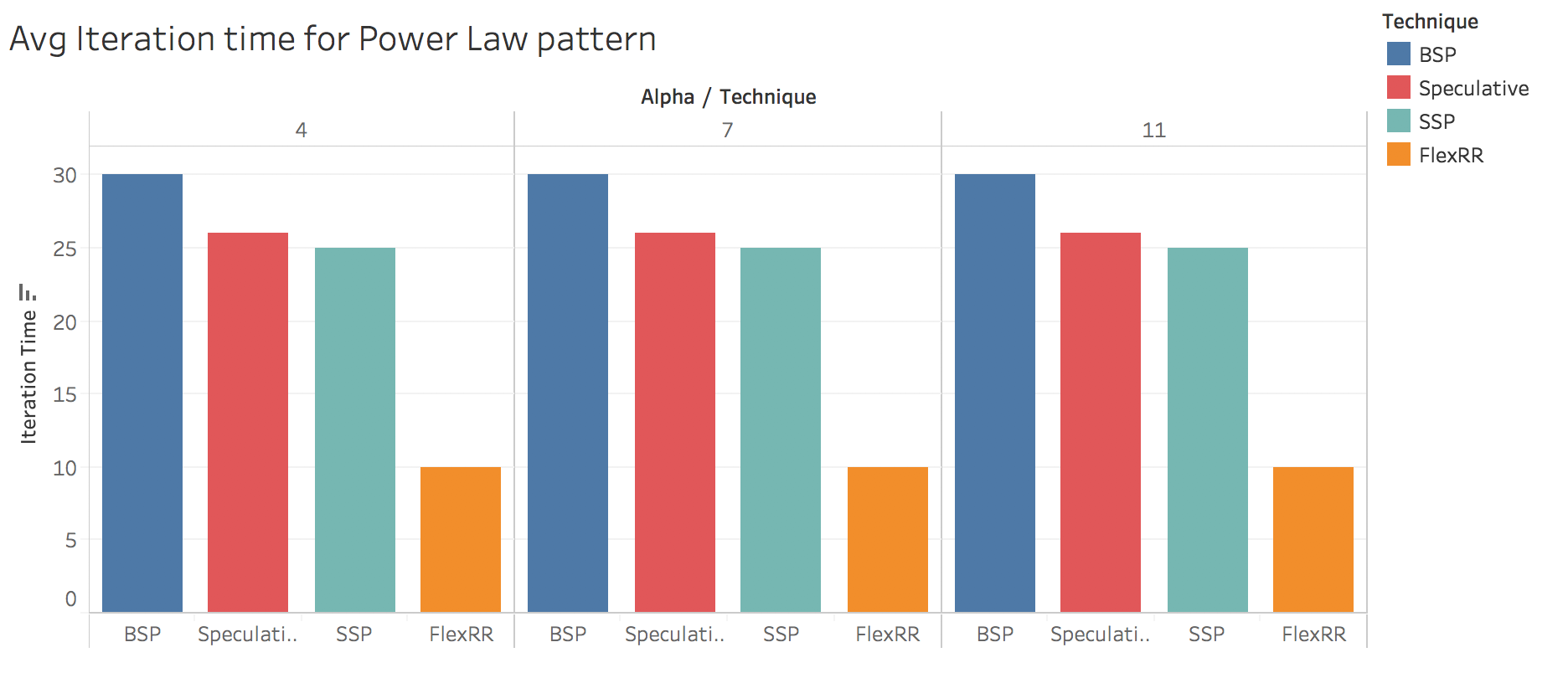}}
\makeatother 
\caption{{Iteration Time for power law machine}}
\label{figure-c2aef6399e14453fb88a0c0a402e88c8}
\end{figure}
\egroup

\bgroup
\fixFloatSize{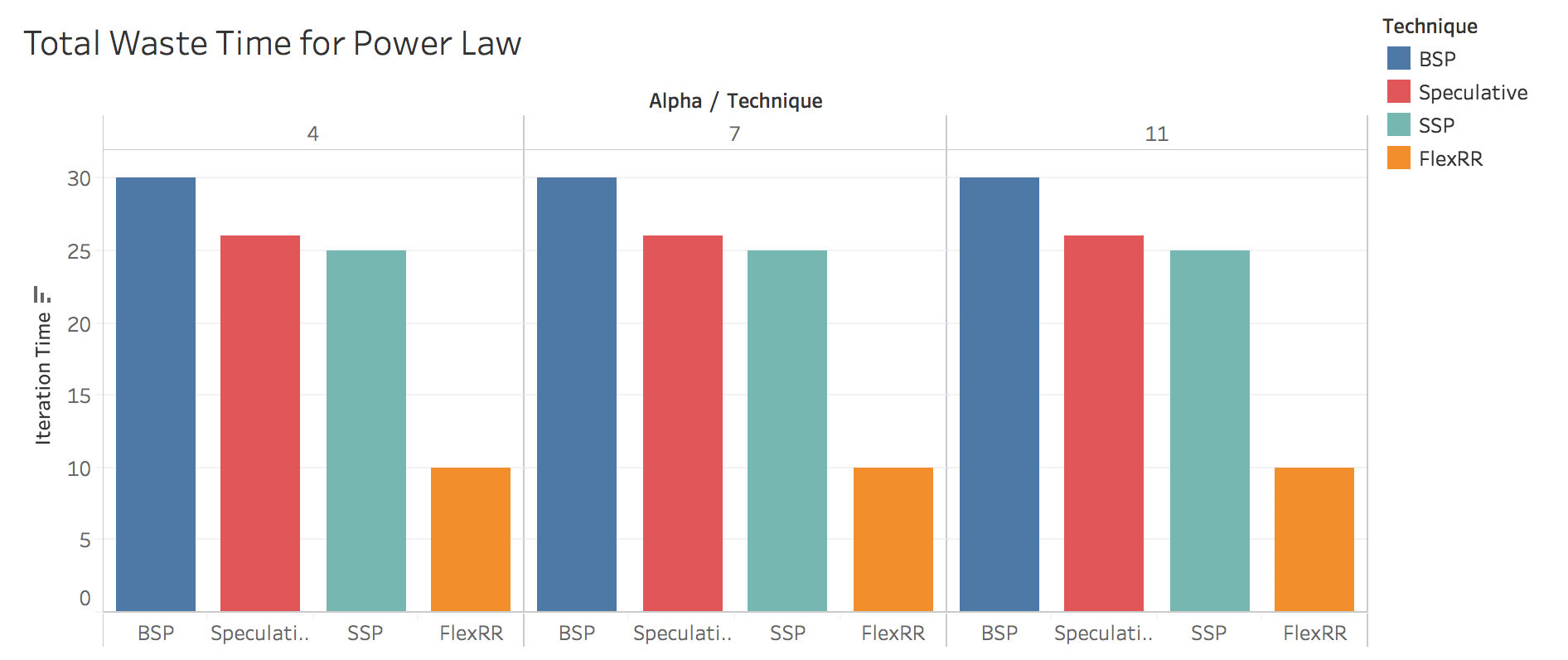}
\begin{figure}[!htbp]
\centering \makeatletter\IfFileExists{images/image8.png}{\includegraphics{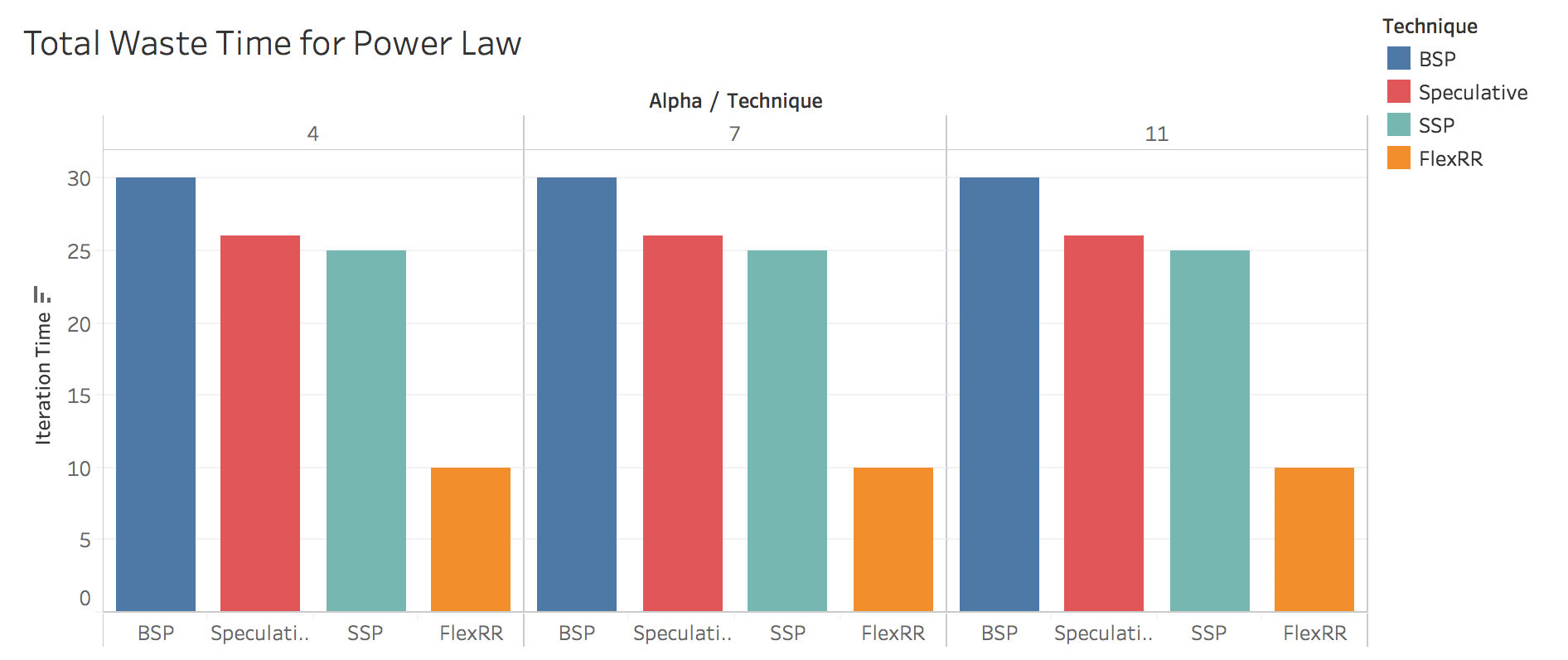}}{\includegraphics{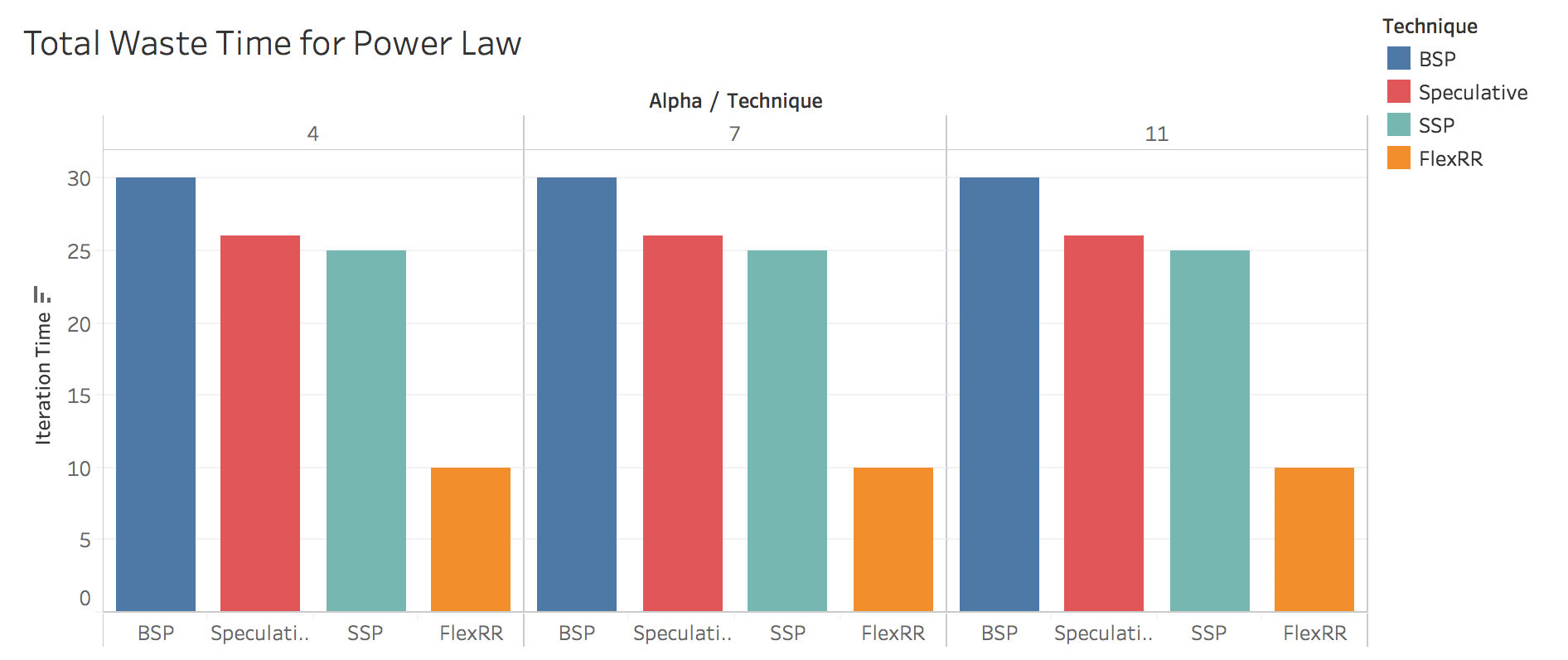}}
\makeatother 
\caption{{Total Waste time for power law machine}}
\label{figure-b0cc6ded711b4332978e7fe086601254}
\end{figure}
\egroup
We compare the average time-per-iteration (20 iterations) and total waste time of alternative modes for the Power-Law Pattern. We present results on each of our applications, setting \ensuremath{\alpha } to 4, 7, and 11. Figure above shows the results for MF. For \ensuremath{\alpha } = 11, SSP and Speculative are faster than BSP by 39\% and 40\%, respectively. When the two techniques are combined in FlexRR, the run-time is 48\% faster than BSP. Similarly, to experiments conducted in earlier sections, with increasing delays (smaller \ensuremath{\alpha }), the other three modes experienced significant increases in run-times, while FlexRR experienced only slight increases.

The results for MLR and LDA show similar trends. For \ensuremath{\alpha } = 11, SSP and BSP RR were 36\% and 31\% respectively faster than BSP for MLR and 37\% and 42\% respectively faster than BSP for LDA. FlexRR was 43\% and 52\% faster than BSP on LR and LDA respectively. With increasing delays (smaller \ensuremath{\alpha }), the other three modes experienced significant increases in run-times for both LR and LDA. FlexRR experienced only modest delays for LR and somewhat larger delays for LDA. In all cases, FlexRR significantly outperforms the other three modes. 
    
\section{Conclusion}
The experiments described herein were designed to address the straggler problem of the Parameter Server for important Machine Learning strategies such as Matrix Factorization (MF), Logistic Regression (LR), and Latent Dirichlet Allocation (LDA), which use iterative approaches to achieve convergence. 

Experiments with custom implemented ML applications under a variety of injected straggler behaviors confirm that FlexRR as a straggler mitigation technique compared to the methods such as the cloning and speculative execution is more effective. The results show that FlexRR decreases the time-per-iteration and wasted time significantly under our experiment environment. The study thus successfully conclude that FlexRR is the best algorithm as a straggler mitigation technique in common cases. 


%

\bibliographystyle{IEEEtran}

\bibliography{article}

\end{document}